\documentclass[%
 aip,
 amsmath,amssymb,
 reprint,%
]{revtex4-1}

\usepackage{amsmath}
\usepackage{graphicx}% Include figure files
\usepackage{dcolumn}% Align table columns on decimal point
\usepackage{bm}% bold math
\usepackage{siunitx}
\usepackage[utf8]{inputenc}
\usepackage[T1]{fontenc}
\usepackage{mathptmx}
\usepackage{color}
\usepackage[english]{babel}
\begin{document}

\preprint{AIP/123-QED}

\title{Reduced variance analysis of molecular dynamics simulations by linear combination of estimators}
% Force line breaks with \\

\author{S. W. Coles}
\affiliation{Department of Chemistry, University of Bath, Claverton Down BA2 7AY, United Kingdom}
\affiliation{The Faraday Institution, Quad One, Harwell Science and Innovation Campus, Didcot OX11 0RA, United Kingdom}

\author{E. Mangaud}
\affiliation{MSME, Universit\'e Gustave Eiffel, UPEC, CNRS, F-77454 Marne-la-Vall\'ee, France}
\author{D. Frenkel}%
\affiliation{
Yusuf Hamied Department of Chemistry, University of Cambridge, Cambridge CB2 1EW, United Kingdom
}%

\author{B. Rotenberg}
\email{benjamin.rotenberg@sorbonne-universite.fr}
% \homepage{http://www.Second.institution.edu/~Charlie.Author.}
\affiliation{%
Physicochimie des \'electrolytes et Nanosyst\`emes Interfaciaux, Sorbonne Universit\'e, CNRS, F-75005 Paris, France
}%

\date{\today}% It is always \today, today,

\begin{abstract}
Building upon recent developments of force-based estimators with a reduced variance for the computation of densities, radial distribution functions or local transport properties from molecular simulations, we show that the variance can be further reduced by considering optimal linear combinations of such estimators. This control variates approach, well known in Statistics and already used in other branches of computational Physics, has been comparatively much less exploited in molecular simulations. We illustrate this idea on the radial distribution function and the one-dimensional density of a bulk and confined Lennard-Jones fluid, where the optimal combination of estimators is determined for each distance or position, respectively. In addition to reducing the variance everywhere at virtually no additional cost, this approach cures an artefact of the initial force-based estimators, namely small but non-zero values of the quantities in regions where they should vanish. Beyond the examples considered here, the present work highlights more generally the underexplored potential of control variates to estimate observables from molecular simulations.
\end{abstract}

\maketitle

The purpose of particle-based atomistic or mesoscopic simulations is to sample phase space in order to compute observables as ensemble averages, with the ultimate goals of predicting physical properties and understanding their microscopic origin. In parallel to the development of more accurate models to describe real systems and of more efficient algorithms and of more powerful computers to generate more data, computational physicists, chemists and biologists always benefitted from advanced statistical tools to design their simulations and analyze their data. Beyond basic results of estimation theory to compute averages and uncertainties, more advanced approaches such as importance sampling have now become standards. While variance-reduction techniques have been around since the early days of molecular simulations, their full potential seems to have been under-explored in this field compared to others (\emph{e.g.} finance) or to other branches of computational Physics such as Quantum Monte Carlo\cite{assaraf_zero-variance_1999,assaraf_improved_2007,toulouse_zero-variance_2007}, or kinetic theory\cite{radtke_variance-reduced_2009,radtke_efficient_2013,peraud_alternative_2012,radtke_low-noise_2011,peraud_efficient_2011}.

Recent promising examples of such endeavours include new strategies for the determination of generic observables~\cite{schultz_reformulation_2016,schultz_alternatives_2019,Purohit2019wzp} and transport properties~\cite{ercole_accurate_2017,baroni_heat_2018,bertossa_theory_2019}, as well as force-based estimators to sample local properties such as number, charge and polarization densities, radial distribution functions (RDF) or local transport properties with a reduced variance~\cite{adib_unbiased_2005,borgis_computation_2013,de_las_heras_better_2018,schultz_alternatives_2019,trokhymchuk_alternative_2019,Purohit2019wzp,Coles2019wca,mangaud_sampling_2020,rotenberg_use_2020}. The availability of different estimators for the same observable opens the possibility of exploiting another well known approach to further reduce the variance, namely the control variates method\cite{hammer_control_2008,boyaval_variance_2010,dellaportas_control_2012,mira_zero_2013,lelievre_partial_2016,oates_control_2017,roussel_perturbative_2019,nguyen_implicit_2020,belomestny_variance_2020}. Here we show the potential of combining of estimators for the determination of RDF and one-dimensional density profiles of a bulk and confined fluid, defined (for a one-component system of $N$ particles in a volume $V$) respectively as the ensemble averages:
\begin{equation}
\label{eq:defg}
g(r) = \frac{V}{N^2} \frac{1}{4\pi r^2}   \left\langle \sum_i \sum_{j\neq i} \delta( r_{ij} - r ) \right\rangle
\end{equation}
with $r_{ij}$ the distance between two particles $i$ and $j$, and
\begin{equation}
\label{eq:defrho}
\rho(z) = \frac{1}{S}\left\langle \sum_i \delta( z_i - z ) \right\rangle    
\end{equation}
with $z_i$ the position of particle $i$ along the $z$-axis and $S$ the area of the system in the lateral direction. 
Before turning to the specific estimators for these observables, let us first introduce the generic idea of combining two estimators $E_0(x)$ and $E_1(x)$ of the same property, which in the present case explicitly depends on some parameter $x$ (distance or position). 
For each value of $x$, one can consider linear combinations of the form:
\begin{equation}
\label{eq:defElambda}
    E_\lambda(x) = (1-\lambda) E_0(x) + \lambda E_1(x) = E_0(x) + \lambda \Delta(x)
\end{equation}
with $\Delta(x)=E_1(x) - E_0(x)$. For any value of $\lambda$, the expectation value $\left\langle E_{\lambda} (x) \right\rangle$ is the same as both $E_0(x)$ and $E_1(x)$; $E_\lambda(x)$ is therefore another valid estimator of the same property. 
However, the variance $\operatorname{var}(E_{\lambda} (x)) = \left\langle \left[ E_{\lambda} (x) - \left\langle E_{\lambda} (x) \right\rangle \right]^2 \right\rangle$
is a quadratic function of $\lambda$. Importantly, one can find for each $x$ the combination that minimizes the variance:
\begin{equation}
\label{eq:optimallambda}
\lambda^*(x)= -\frac{\operatorname{cov}(E_0(x),\Delta(x))}{\operatorname{var}(\Delta(x))}= 1 -\frac{\operatorname{cov}(E_1(x),\Delta(x))}{\operatorname{var}(\Delta(x))} \; ,
\end{equation}
which involves covariances $\operatorname{cov}(A,B) = \langle (A - \langle A \rangle) (B - \langle B \rangle)\rangle$. This provides for each $x$ the optimal estimator $E_{\lambda}^*(x)$ in this family, 
which has a reduced variance compared to both $E_0(x)$ and $E_1(x)$:
\begin{equation}
\label{eq:varoptElambda}
\operatorname{var}(E_{\lambda}^* (x)) = 
\left[1-\rho_{E_n,\Delta}^2(x)\right]\operatorname{var}(E_n(x))
\end{equation}
with $\rho_{E_n,\Delta}(x)=\operatorname{cov}(E_n,\Delta)/\sqrt{\operatorname{var}(E_n)\operatorname{var}(\Delta)}$ the correlation coefficient between $E_n(x)$ and $\Delta(x)=E_1(x)-E_0(x)$. In practice, given two estimators and a set of configurations, one simply computes for each $x$ both expectation values $\left\langle E_n(x)\right\rangle$ as well as the (co-)variances entering in Eq.~\ref{eq:optimallambda} to obtain the corresponding optimal estimator $E_{\lambda}^* (x)$ (see Eq.~\ref{eq:defElambda}).

We now introduce pairs of estimators for the RDF (Eq.~\ref{eq:defg}) and the one-dimensional density distribution (Eq.~\ref{eq:defrho}). In each case, both estimators use the force acting on the atoms in addition to their position and correspond to an ``integral of the gradient'', but they differ in the choice of the origin to perform this integration. For the RDF, we first consider the estimator introduced in Ref.~\citenum{borgis_computation_2013}, which uses the limit value of 1 for $r\to\infty$, $g_\infty(r)=\left\langle \hat{g}_{\lambda=0}(r) \right\rangle$ with
\begin{equation}
\label{eq:defglambda0}
\hat{g}_{\lambda=0}(r) = 1+\frac{V}{N^2} \frac{\beta}{4\pi} \sum_i \sum_{j\neq i} 
\frac { \left( \mathbf { f } _{ j } - \mathbf { f }_ { i } \right)} { 2 }  \cdot \frac { \mathbf { r } _{ i j } } { r_ { i j } ^ { 3 } } H \left( r _ { i j } -r \right)
\; ,
\end{equation} 
where $\mathbf { f }_ { i }$ and $\mathbf { f }_ { j }$ are the forces acting on atoms $i$ and $j$, $\mathbf { r }_ { ij }$ is the displacement between atoms $i$ and $j$, $\beta=1/k_BT$ with $k_B$ the Boltzmann constant and $T$ the temperature, and $H$ is the Heaviside function. Compared to the standard histogram approach, obtained by discretizing the definition Eq.~\ref{eq:defg}, in this force-based estimator all pairs separated by a distance larger than $r$ contribute to the estimate of the RDF at $r$, which reduces the variance considerably. In addition, no bins are necessary and the RDF can be obtained with arbitrary resolution. However, Eq.~\ref{eq:defglambda0} leads to a small yet spurious non-vanishing value (and non-zero variance) for $r\to0$. An alternative expression was proposed in Ref.~\citenum{Purohit2019wzp} using this condition inside the core and we therefore also consider $g_0(r)=\left\langle \hat{g}_{\lambda=1}(r)\right\rangle$, with
\begin{equation}
\label{eq:defglambda1}
\hat{g}_{\lambda=1}(r) = \frac{V}{N^2} \frac{\beta}{4\pi} \sum_i \sum_{j\neq i} 
\frac { \left( \mathbf { f } _{ j } - \mathbf { f }_ { i } \right)} { 2 }  \cdot \frac { \mathbf { r } _{ i j } } { r_ { i j } ^ { 3 } } H \left( r - r _ { i j } \right)
\; ,
\end{equation} 
where this time all pairs separated by a distance smaller than $r$ contribute to the estimate of the RDF at $r$. Symmetrically to $g_\infty(r)$, $g_0(r)$ displays by construction a vanishing value with zero variance inside the core, but a larger variance at large distance. Note that the subscripts in  $g_\infty$ and $g_0$ refer to the distance from which the gradient is integrated, not to a value of the mixing parameter $\lambda$.

The gradient of the 1D-density is given by the force density $\left\langle\frac{1}{S} \sum_i\delta( z - z_i ) \beta f_{z,i} \right\rangle$. For a fluid confined between walls, the density can therefore be obtained by integrating the force density from either side, starting from 0 inside the walls. This leads to two complementary estimators defined by:
\begin{equation}
\label{eq:defrho0}
\rho_0(z)= \left\langle \hat{\rho}_{\lambda=0}(z) \right\rangle = \left\langle\frac{1}{S} \sum_i H( z - z_i ) \beta f_{z,i}\right\rangle
\; ,
\end{equation}
and 
\begin{equation}
\label{eq:defrhoL}
\rho_L(z)=\left\langle \hat{\rho}_{\lambda=1}(z) \right\rangle = \left\langle\frac{1}{S} \sum_i H( z_i - z ) \beta f_{z,i}\right\rangle
\; .
\end{equation}
Both estimators identically vanish (and so does the corresponding variance) on one side but have a spurious non-zero value (and corresponding variance) on the other.

We illustrate the advantage of combining estimators for a simple Lennard-Jones (LJ) fluid at a reduced density $\rho^*=\rho\sigma^3=0.8$ and reduced temperature $T^*=k_BT/\epsilon=1.35$, with $\sigma$ and $\epsilon$ the LJ diameter and energy, respectively. For the RDF we consider a bulk system with $N=864$ particles, while for the 1D-density we simulate a fluid with $N=1152$ particles confined between two walls, consisting each of $72$ LJ particles identical to the fluid on a centered square lattice (with lattice spacing $\sqrt{2}\sigma$) and separated by a distance $22\sigma$. We use periodic boundary conditions in 3 and 2 dimensions for the bulk and confined systems, respectively. In both cases we use a time step of $10^{-3}~t^*$ with $t^*=\sqrt{m\sigma^2/\epsilon}$ the LJ time unit and $m$ the mass of the particles, and a Nos\'e-Hoover thermostat with a time constant of $0.1~t^*$. 
In order to estimate the relevant averages, (co-)variances and resulting optimal combination parameter $\lambda^*$, we use $10^3$ configurations separated by $t^*$ in both the bulk and confined cases. Even though the force-based estimators do not require bins and can be evaluated at arbitrary positions, the results are shown for evenly spaced distances and positions with $\Delta r=0.005\sigma$ and $\Delta z = 0.005\sigma$ for the RDF and 1D-density, respectively.

\begin{figure*}[ht!]
\centering
\includegraphics[width=15cm]{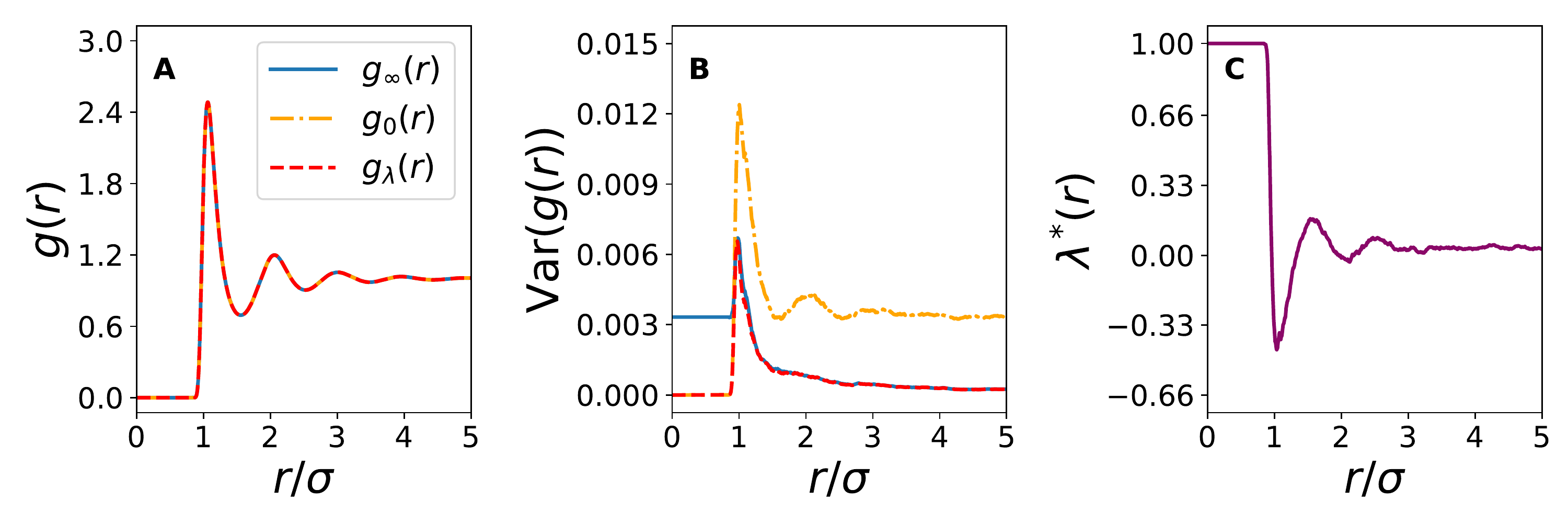}
\caption{
(a) Radial distribution function for a bulk Lennard-Jones fluid with the two estimators $g_\infty$ (see Eq.~\ref{eq:defglambda0}, solid blue line) and $g_0$ (see Eq.~\ref{eq:defglambda1}, dashed-dotted yellow line), corresponding to $\lambda=0$ and $\lambda=1$, respectively, and with the optimal linear combination for each distance ($g_\lambda$, see Eqs.~\ref{eq:defElambda} and~\ref{eq:optimallambda}, dashed red line). (b) Variance with each estimator. (c) Optimal mixing parameter $\lambda^*$ as a function of distance. In all panels the distance is normalized by the LJ diameter $\sigma$.
}
\label{fig:RDF}
\end{figure*}

\begin{figure*}[ht!]
\centering
\includegraphics[width=15cm]{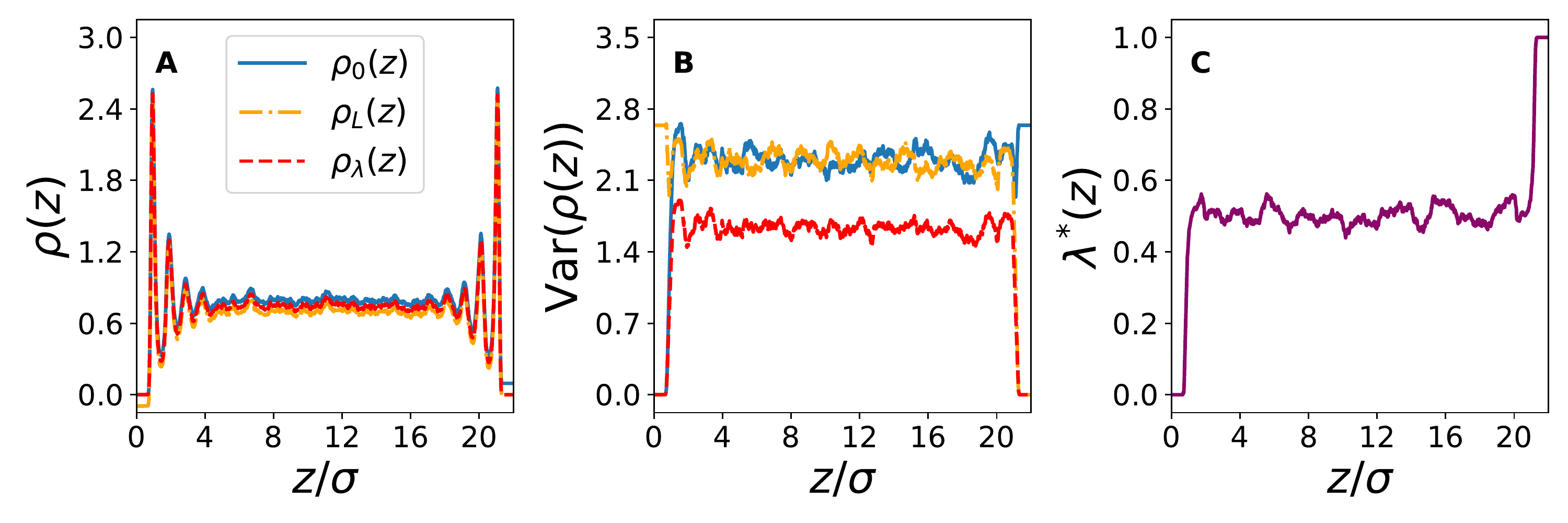}
\caption{
(a) One-dimensional density profile for a confined Lennard-Jones fluid, estimated with the two estimators $\rho_0$ (see Eq.~\ref{eq:defrho0}, solid blue line) and $\rho_L$ (see Eq.~\ref{eq:defrhoL}, dashed-dotted yellow line), corresponding to $\lambda=0$ and $\lambda=1$, respectively,
and with the optimal linear combination for each distance ($\rho_\lambda$, see Eqs.~\ref{eq:defElambda} and~\ref{eq:optimallambda}, dashed red line). (b) Variance divided by the density with each estimator. (c) Optimal mixing parameter $\lambda^*$ as a function of position. In all panels the position is normalized by the LJ diameter $\sigma$.
}
\label{fig:1ddensity}
\end{figure*}

Results for the RDF and the 1D-density profile are shown in Figures~\ref{fig:RDF} and~\ref{fig:1ddensity}, respectively. In both cases all estimators provide the same results for the observables (panels \ref{fig:RDF}a and~\ref{fig:1ddensity}a), which are typical of bulk and interfacial fluids. However the variances (panels \ref{fig:RDF}b and~\ref{fig:1ddensity}b) differ.  The variance of the optimal combination is lower than that of both initial estimators, as expected. It is also instructive to consider the optimal weight $\lambda^*$, shown in panels \ref{fig:RDF}c and~\ref{fig:1ddensity}c. In both cases, the optimal combination coincides with the zero-variance estimator in the corresponding regions: inside the core for $g_0$ and at large distance for $g_\infty$, for $z\to 0$ and $z\to L$ for $\rho_0$ and $\rho_L$, respectively. Beyond these limits, the contributions of the two estimators are equal in the central region for the 1D-density, but their evolution is more complex for the RDF, with oscillations following that of the RDF itself and even a negative region for $\lambda^*$ near the first peak of $g(r)$. 

Beyond the illustration on the RDF and 1D-density, this work highlights the potential of the control variates method to combine estimators of the same quantity in order to obtain new estimators with reduced variance. This approach is particularly beneficial when the regions in which the initial estimators perform well in complementary regions, as it mitigates their respective limitations. The additional cost is limited, since it only requires computing (co-)variances of the initial estimators. The recent development of alternative estimators for molecular simulations, \emph{e.g.} force-based or within the mapped-averaging framework, can directly benefit from the present control variates approach, for other observables such as angular distributions or two-particle densities. One could also consider other combinations, such as direct estimates of response functions (\emph{e.g.} heat capacity or capacitance) with their fluctuation counterparts (\emph{e.g.} energy or charge fluctuations). 

\section*{Acknowledgements}

The authors thank Gabriel Stoltz and Tony Leli\`evre for useful suggestions. 
This project has received funding from the European Union’s Horizon 2020 research and innovation programme under grant agreement No. 766972 and from the European Research Council under the European Union's Horizon 2020 research and innovation programme (grant agreement No. 863473). S.~W.~C.\ acknowledges the support of the Faraday Institution through the CATMAT project (grant number FIRG016) and the Balena High Performance Computing Service at the University of Bath.

\section*{Data availability}

The data that support the findings of this study are available from the corresponding author upon reasonable request.

\bibliography{refs}% Produces the bibliography via BibTeX.

\end{document}